# First-principles Study of the Interactions of Electron Donor and Acceptor Molecules with Phosphorene

Ruiqi Zhang,[a] Bin Li,[a,b] and Jinlong Yang[a, b, *]

[a] Hefei National Laboratory for Physics at Microscale, University of Science and Technology of China, Hefei, Anhui 230026, China

[b] Synergetic Innovation Center of Quantum Information & Quantum Physics, University of Science and Technology of China, Hefei, Anhui 230026, China

**ABSTRACT**

Density functional theory calculations have been carried out to investigate single-layer phosphorene functionalized with two kinds of organic molecules, i.e. an electrophilic molecule tetracyano-p-quinodimethane (TCNQ) as electron acceptor and a nucleophilic molecule tetrathia-fulvalene (TTF) as electron donor. The TCNQ molecule introduces shallow acceptor states in the gap of phosphorene close to the valence band edge (VBE), which makes the doped system a p-type semiconductor. However, when the TTF molecule is adsorbed on the phosphorene, the occupied molecular states introduced into the gap are of deep donor states so that effective n-doping for transport cannot be realized. This disadvantageous situation can be amended by applying an external electric field perpendicular to the phosphorene surface with direction from the phosphorene to the TTF molecule, under which the TTF-introduced donor states move closer to conduction band edge (CBE) of the phosphorene and then the TTF-doped phosphorene system becomes an n-type semiconductor. The effective bipolar doping of single-layer phosphorene via molecular adsorption predicted above, especially n-doping against its native p-doping propensity, would broaden the way to the application of this new type of two-dimensional material in nanoelectronic and optoelectronic devices.

# I. INTRODUCTION

In recent years, graphene-like two-dimensional (2D) materials, owing to their remarkable properties,[1-2] have attracted much research attention as emerging device materials for nanoelectronics. These materials, such as graphene,[3-4] silicene,[5-7] boron-nitride nanosheets,[8-10] and 2D transition-metal dichalcogenides (TMDs),[11] have novel properties which are different from their bulk counterparts. With many promising applications in nanoelectronics and optoelectronics, the 2D materials are consider to be a relatively new and exciting area for nanotechnology.

Recently, a new 2D semiconducting material, namely, few-layer black phosphorus or phosphorene, has been isolated successfully by mechanical exfoliation from black phosphorus crystals.[12-13] Unlike the widely studied other 2D materials, phosphorene exhibits a sizable and novel direct band gap, which can be modified from 1.51 eV for a monolayer to 0.59 eV for a five-layer.[14] Besides, the phosphorene-based FET exhibits high mobility of 286 $cm^2$/V/s and phosphorene possesses highly on/off ratios, up to $10^4$ mobility.[10] With some other interesting and useful features, such as its anisotropic electric conductance and optical responses, phosphorene is considered to be a remarkable 2D material for electronic and optical applications[15] and has immediately received considerable research.[16-18]

On the basis of our current understanding, well-behaved p-type FETs based on few-layer phosphorene[12,14] and black phosphorus-monolayer $MoS_2$ van der Waals heterojunction p-n diode were fabricated in experiment.[19] The p-type conductivity observed in these experiments of the phosphorene sample was attributed to the intrinsic point defect states of the phosphorene.[20] It is well-known that many electronic devices, such as p−n junctions and complementary metal-oxide-semiconductors (CMOS), require the ability of the semiconductor to be both p-doped and n-doped. Especially, the effective n-type doping of the phosphorene is important because of its native p-type doping property. In this paper, we discuss the possibility of obtaining effectively p-doped and n-doped single-layer phosphorene by employing method of molecular doping.

The molecular doping is a promising, simple and effective way to tune the electronic structures of 2D materials via their interactions with the molecule, because a large variety of inorganic and organic molecules are available. Especially, a lot of the organic molecules have been designed and synthesized, and different functional groups can be incorporated into them and their structures still keep stable. Owing to these, organic molecules have some advantages in surface doping. For emerging nanomaterials such as carbon nanotubes,[21] graphene, boron nitride nanotubes,[22-23] hexagonal boron nitride nanosheeets,[24] $MoS_2$[25] and so on, tremendous theoretical and experimental efforts have been devoted to study the interactions of these nanomaterials with different organic molecules. To our best knowledge, there is no theoretical or experimental report on the noncovalent surface doping via the organic molecules on phosphorene. Here, by employing reliable first-principles calculations, we carry out a systematic study to explore the interactions between the single-layer phosphorene and two typical organic molecules with aim of realizing effective n- and p-doping of phosphorene. As we know, a typical p-type dopant should have empty states close to the VBE, and on the contrary, a typical n-type dopant should have filled states close to the CBE. In this paper two kinds of organic molecules are used to dope phosphorene: one is electrophilic molecules TCNQ possessing effective electron-withdrawing capabilities with a electron affinity (EA) of 2.80 eV; the other is nucleophilic molecules TTF possessing effective electron-donating capabilities with a ionization potential (IP) of 7.36 eV. The structures of these two molecules are shown in Figs. 1(a) and 1(b). Our results indicate that phosphorene is still a p-type semiconductor after adsorbing TCNQ molecule as acceptor. On the other hand, after adsorbing TTF molecule as donor there appear molecule-derived filled states in the band gap of phosphorene but these states are still far from the CBE. It is also found that if an appropriate external electric field vertical to the phosphorene surface is applied, the effective n-doping of phosphorene will be realized.

## II. THEORETICAL MODEL AND METHOD

In this work, our first-principles calculations are based on the density functional theory (DFT) implemented in the VASP package.[26] The generalized gradient approximation of Perdew, Burke, and Ernzerhof (GGA-PBE) and projector augmented wave (PAW) potentials are used.[27] In all computations, the kinetic energy cutoff are set to be 520 eV in the plane-wave expansion. The Brillouin zone is sampled with a grid of 3×3×1 conducted by the Monkhorst-Pack special k-point scheme.[28] All the geometry structures are fully relaxed until energy and forces are converged to $10^{-5}$ eV and 0.02 eV/Å, respectively. Dipole correction is employed to cancel the errors of electrostatic potential, atomic forces and total energy, caused by periodic boundary condition.[29] Effect of van der Waals (vdW) interaction is accounted for by using the empirical correction method proposed by Grimme (DFT-D2) [30], which is a good description of long-range vdW interactions.[31-34] As a benchmark, DFT-D2 calculations give a interlayer distance of 3.25 Å and a binding energy of -25 meV per carbon atom for bilayer graphene, consistent with previous experimental measurements and theoretical studies.[35-36] In order to estimate the charge transfer between the organic molecules and phosphorene, we adopt Bader charge analysis method,[37] using the Bader program of Henkelman's group.[38-40]

Unlike a flat structure of graphene, the single-layer phosphorene has a puckered honeycomb structure with each phosphorus atom covalently bonded with three adjacent atoms (Fig. 1(c)). The initial structure of phosphorene used in our simulation is obtained from black phosphorus.[41] Our calculated lattice constants for bulk black phosphorus are a = 3.304 Å, b = 4.561 Å, and c = 11.307 Å, in good agreement with experimental values and other theoretical calculations.[42] The relaxed lattice constants for the single-layer phosphorene are a = 3.298 Å, b = 4.624 Å. To study the adsorption of TCNQ and TTF molecule on the single-layer phosphorene, we adopt a 4×6×1 supercell containing 96 P atoms, so the molecule-phosphorene adsorption system consists of a 18.49 Å × 19.79 Å × 20.18 Å supercell (Fig. 1(d)). The nearest distance between two TCNQ/TTF molecules in adjacent supercells is ~ 10 Å, and a large value 15 Å of the vacuum region is used to avoid interaction between two adjacent periodic images. In order to evaluate stability of the molecular adsorption on phosphorene, adsorption energy is defined as

$$E_a = E_{Molecule+Phosphorene} - E_{Molecule} - E_{Phosphorene} \quad (1)$$

where $E_{Molecule+Phosphorene}$, $E_{Molecule}$ and $E_{Phosphorene}$ represent total energy of molecular adsorption on phosphorene, molecules and phosphorene, respectively.

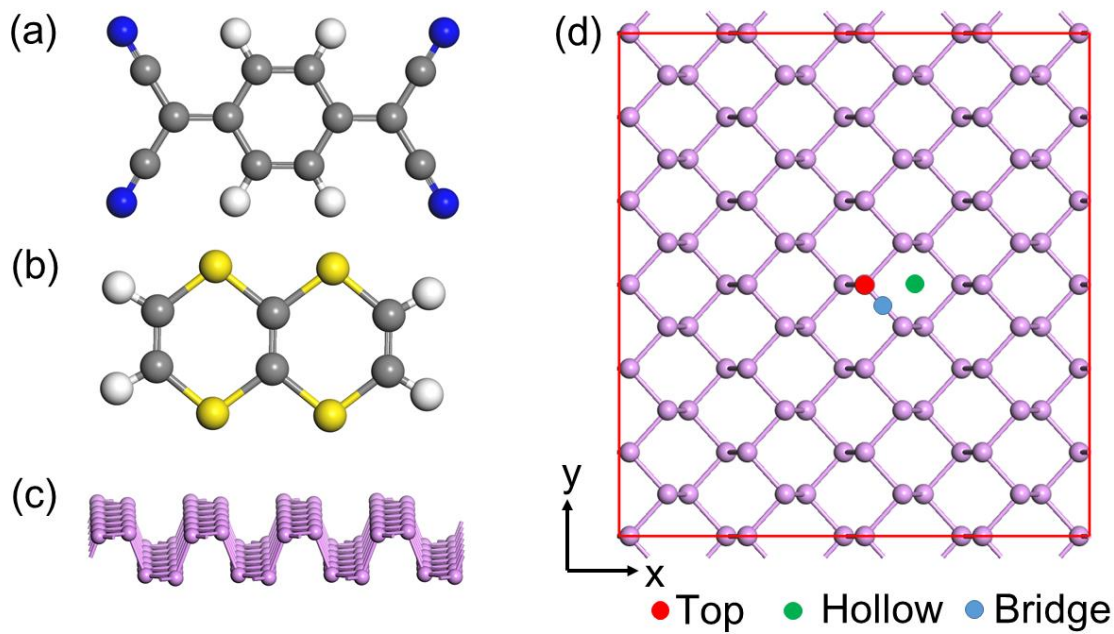

**Figure 1.** (a) Top view of molecular structure for TCNQ. (b) Top view of molecular structure for TTF. (c) Side view for single-layer phosphorene. (d) Top view of the $4\times6\times1$ supercell of phosphorene, with three possible molecular adsorption sites being marked by different color circular.

## III. RESULTS AND DISCUSSION

To understand how the molecules are adsorbed on phosphorene, we firstly investigate which adsorption site is the most stable one. Due to the highly symmetrical geometries of TTF and TCNQ molecules, three adsorption sites are considered for each molecule (see Fig. 1(d)): the top of a phosphorus atom (T), the center of a phosphorus hexagon (H), and the center of a phosphorus-phosphorus bond (B). For each of these positions, two molecular orientations are tested, that is to say, the long axis of the organic molecules being along the x-direction or y-direction of the unit cell. The adsorbed molecules are initially placed parallel to the phosphorene surface with a vertical distance of 2.7 Å. We have considered various initial configurations for the molecular adsorption on phosphorene, as mentioned above. The adsorption energy for different adsorption configurations after relaxation are shown in Table. 1, from which we learned that the adsorption energy difference between different configurations is small.

**Table.1** The adsorption energies ($E_a$) of TCNQ and TTF on phosphorene for different adsorption configurations.

| System | Position | Orientation | $E_a$ (eV) |
|--------|----------|-------------|------------|
| TCNQ   | T        | x           | -1.388     |
|        |          | y           | -1.368     |
|        | H        | x           | -1.142     |
|        |          | y           | -0.741     |
|        | B        | x           | -1.254     |
|        |          | y           | -1.319     |
| TTF    | T        | x           | -1.510     |
|        |          | y           | -1.283     |
|        | H        | x           | -1.510     |
|        |          | y           | -1.558     |
|        | B        | x           | -1.514     |
|        |          | y           | -1.460     |

Next, we only make further detailed investigations on the most stable configurations. The most stable configuration for TCNQ is the top site, with its long axis along x-direction of unit cell (see Fig. 2(a)). While, for TTF, the most stable configuration is the hollow site with its long axis along y-direction of unit cell (see Fig. 2(b)). Here we define the equilibrium distance between the molecule and phosphorene as the distance between the nearest atoms of two subsystems. The equilibrium distance between TCNQ and phosphorene is 3.10 Å, meanwhile both TCNQ and phosphorene are slightly crumpled from the flat plane due to weak interaction. That is to say, the interaction between the TCNQ molecule and phosphorene is mainly of physisorption. In comparison with the case of TCNQ, the equilibrium distance between the TTF molecule and phosphorene is 2.47 Å. The adsorbed TTF molecule is severely bent, and the maximal height difference of the atoms in the molecule reaches 1.14 Å (see Fig. 2(b)). The adsorption of TTF also results in a slight distortion of phosphorene.

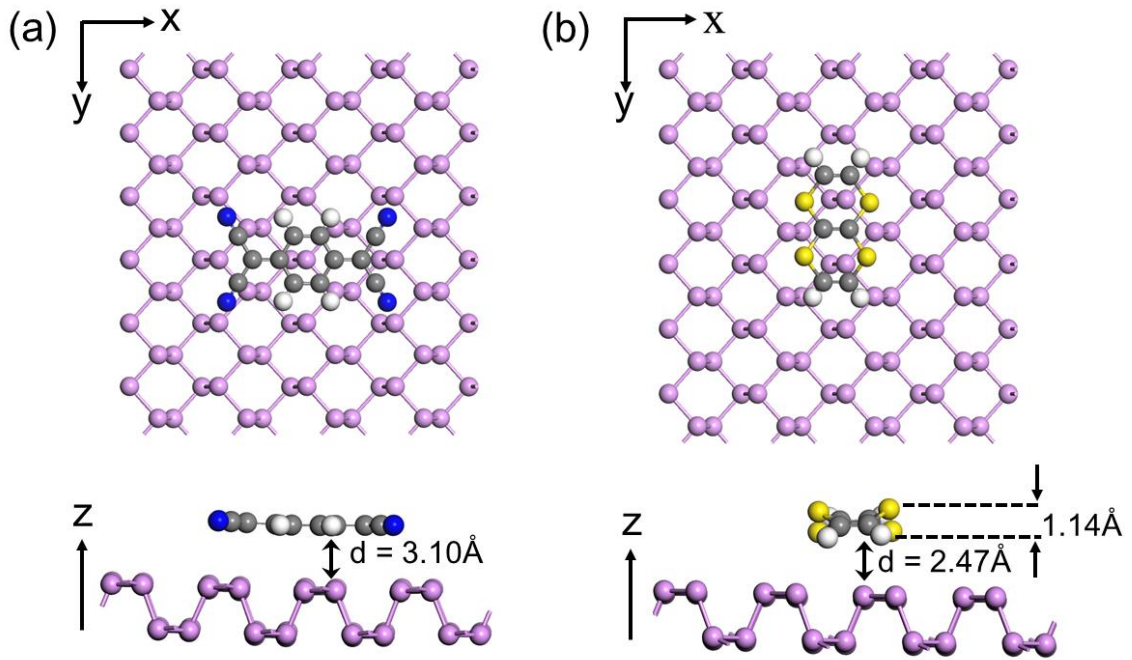

**Figure 2.** Top and side views of the most stable configurations for the adsorbed (a)TCNQ and (b) TTF molecules on phosphorene, and the equilibrium distances between the phosphorene and the molecules as well as the bending distance of the TTF molecule are also given

In order to examine how these adsorbed molecules affect the electronic properties of phosphorene, we calculate the electronic band structures of single-layer phosphorene before and after the molecular adsorption. The electronic band structures of pristine phosphorene, TCNQ-adsorbed phosphorene system and TTF-adsorbed phosphorene system are shown in Figs. 3(a)-3(c). Clearly, the pristine phosphorene is a direct-gap semiconductor with a band gap of 0.91 eV at Γ point calculated using PBE, in good agreement with previous report.[11] Comparing Fig. 3(a) with Fig. 3(b), we find that a flat band appears just above the Fermi level ($E_F$) after the adsorption of TCNQ, and it is guessed to be mostly contributed by the TCNQ, named as the lowest unoccupied molecular band. However, for the TTF-adsorbed phosphorene system, a flat band guessed to be mainly derived from the TTF appears just below the $E_F$, named as the highest occupied molecular band. It is obvious that these two bands may act as the acceptor state in the TCNQ-adsorbed phosphorene system and donor state in the TTF-adsorbed phosphorene system, respectively.

The appearances of the acceptor and donor state due to the molecular adsorption decease the band gap of phosphorene. We define the new band gap between the lowest unoccupied molecular band and the VBE of phosphorene as $E_p$, and the new band gap between the highest occupied molecular band and the CBE as $E_n$. A typical p-doped (n-doped) semiconductor requests the enough small $E_p$ ($E_n$), which means lower activation energy for the holes (electrons) localized at impurity to become conductive carriers. For the TCNQ-adsorbed phosphorene system, the lowest unoccupied molecular band (acceptor states) become very close to the phosphorene VBE and $E_p$ is only 0.12 eV, showing that it forms the shallow acceptor states, so a typical p-type doping is realized and there maybe exists considerable charge transfer between the TCNQ molecule and phosphorene. In fact, according to our Bader charge analysis, the charge transfer between them is about 0.35 electron per molecule from phosphorene to TCNQ. Such considerable electron transfer also indicates that the TCNQ-adsorbed phosphorene system is a typical p-type semiconductor. For the TTF-adsorbed phosphorene system, the highest occupied molecular band (donor states) is found to be also close to the phosphorene VBE, not CBE, and $E_n$ reaches 0.73 eV, which is too large to form an effective n-type doping. In fact, it belongs to the case of deep doping and the highest occupied molecular band forms the deep donor states. In addition, Bader charge analysis shows that there is only 0.06 electron per molecule transferring from TTF to phosphorene. Such small charge transfer also suggests that the TTF-adsorbed phosphorene system is not a typical n-type semiconductor.

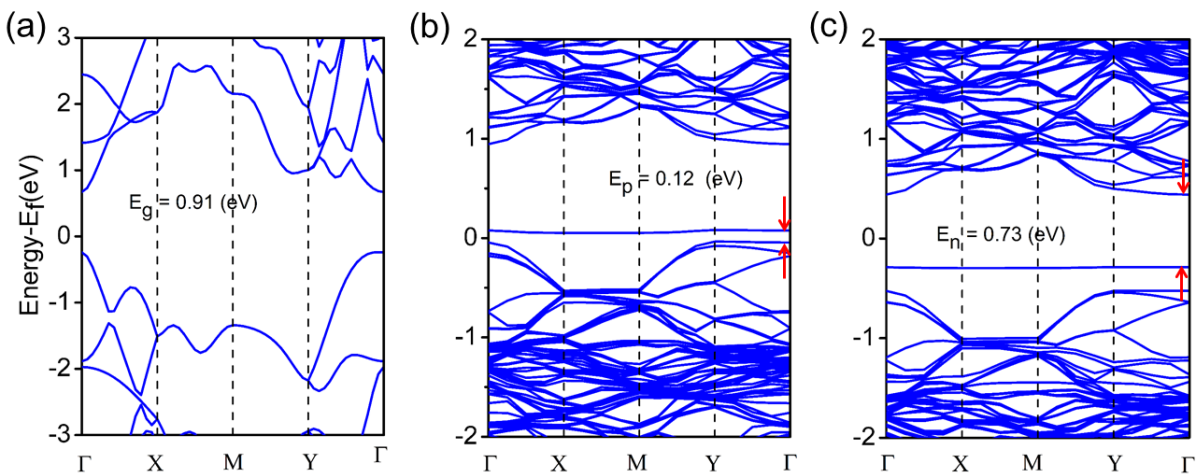

**Figure 3**. Calculated band structures of (a) pristine phosphorene, (b) TCNQ-adsorbed phosphorene, and (c) TTF-adsorbed phosphorene ; Γ (0.0, 0.0, 0.0), X (0.0, 0.5, 0.0), M (0.5, 0.5, 0.0), and Y (0.5, 0.0, 0.0) refer to special points in the first Brillouin zone.

In order to validate the above conclusions about the doping of phosphorene, we calculate the density of states (DOS) for TCNQ/TTF functionalized phosphorene and the projected density of states (PDOS) for phosphorene and TCNQ/TTF in two adsorption systems (Fig. 4). As shown in Fig. 4(a), for the TCNQ-adsorbed phosphorene system, the introduced new state above the $E_F$, which is very close to the VBE of phosphorene, is indeed mainly ascribed to the TCNQ molecule. And for the TTF-adsorbed phosphorene system, a new state is introduced below the $E_F$, far away from the CBE of phosphorene, and it is TTF molecule that provides the main contribution to this state (see Fig. 4(b)). These results are consistent with the above conclusions that the TCNQ-adsorbed phosphorene system is a typical p-type semiconductor and the TTF-adsorbed phosphorene system is not a typical n-type semiconductor.

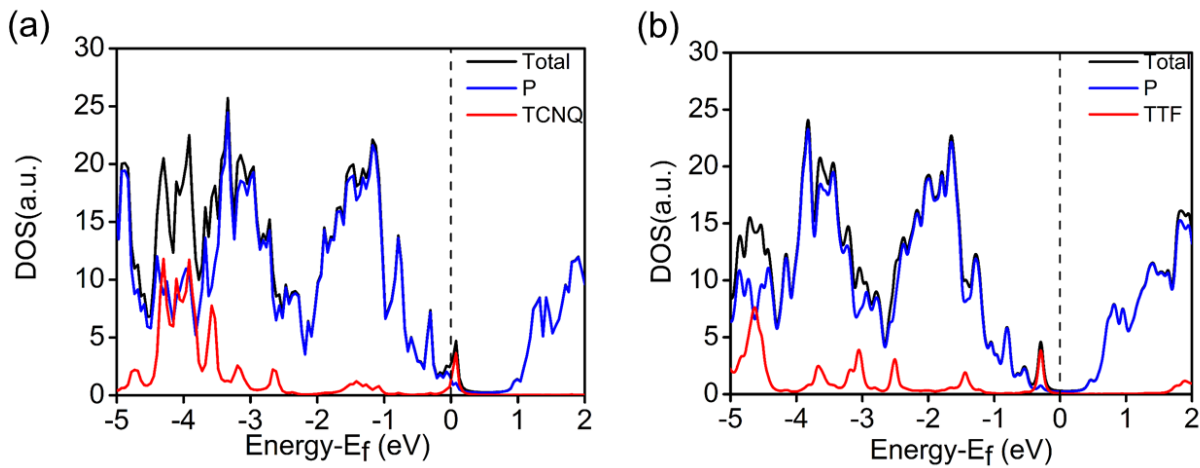

**Figure 4**. Total DOS and PDOS of the molecule and phosphorene for (a) the TCNQ-adsorbed and (b) TTF-adsorbed phosphorenes. The dashed line indicates the $E_F$.

The CBE of phosphorene is ~ 4.10 eV below the vacuum energy level, which means that introducing a dopant with an IP close to 4.10 eV may form a typical n-type semiconducting phosphorene. As mentioned above the IP of TTF is 7.36 eV, so the TTF-adsorbed phosphorene system does not become a typical n-type semiconductor. In order to realize the effective n-doping of phosphorene and obtain a typical n-type semiconducting phosphorene, next we try to explore possible modulation of the band structure of the TTF-adsorbed phosphorene under a varied external electric field. Previous studies had showed that the external electric field may affect the charge transfer and the energy level alignment between the adsorbed species and substrate,[22, 43-44] so we expect that the TTF-adsorbed phosphorene can turn to a typical n-type semiconductor under an appropriate external electric field. The electric field is applied perpendicular to the phosphorene surface, with a strength ranging from -0.5 V/Å to 0.5 V/Å.

Here we define the electric field pointing downward to the phosphorene surface as positive (see red arrow in Fig.5 (a)). As illustrated in Fig. 5(a), $E_n$ of the TTF-adsorbed system varies from 0.22 eV under the electric field of -0.5 V/Å to 0.91 eV under the electric field of 0.5 V/Å based on our PBE calculation, showing a fine controllability. It is observed the highest occupied molecular band moves close to the phosphorene CBE under a negative electric field. Especially, this band becomes the shallow donor state and the TTF-adsorbed phosphorene system is converted to a typical n-type semiconductor at -0.5 V/Å. On the other hand, the highest occupied molecular band gets closer to and even merges into the phosphorene VBE with increasing electric field. Band structures of the TTF-adsorbed phosphorene system under the electric fields of -0.5 V/Å and 0.5 V/Å are plotted in Fig. 5(c) and Fig. 5(d), respectively. We have also calculated the charge transfers between the TTF molecule and phosphorene under different electric fields (also see Fig. 5(a)). Applying the negative electric field is found to enhance the charge transfer. When the field reaches -0.5 V/Å, there is about 0.26 electron transferring from the TTF molecule to phosphorene, also suggesting the effective n-doping of phosphorene.

At last, this modulation of band structure of the TTF-adsorbed phosphorene system under the external electric field can be understood by examining its electrostatic potential profile. The xy-averaged electrostatic potentials of this system under different external electric fields are plotted in Fig. 5(b). It is revealed that the electrostatic potential below the phosphorene surface increases with increasing electric field, but the electrostatic potential of TTF decreases as the electric field increases, so the electric potential difference between them increase when the electric field changes from the negative to positive. Obviously, if we apply a negative electric field on the TTF-adsorbed phosphorene surface, electrons transfer more easily from TTF to phosphorene due to the increased electrostatic potential difference of TTF relative to phosphorene, i.e. better electron donation to phosphorene. At the same time, this increase of the electrostatic potential difference also suggests that the highest occupied molecular band will move upward relative to the CBE of phosphorene, and then turn from the deep donor states to the shallow donor states. So the TTF-adsorbed phosphorene is converted to an effectively n-doped semiconductor under an appropriate negative external electric field.

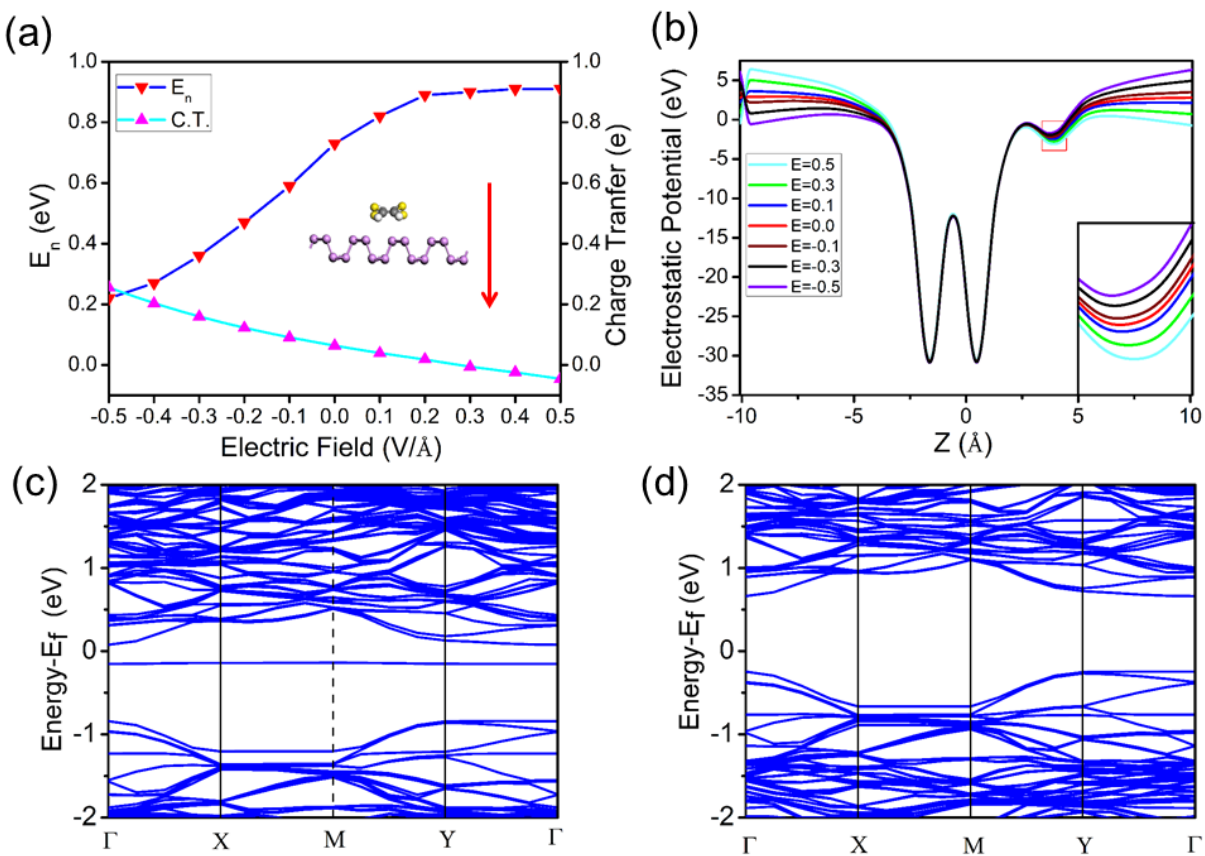

**Figure 5**. (a) Variations of $E_n$ and charge transfer of the TTF-adsorbed phosphorene system under the external field, and the positive direction of electric field is denoted by the red arrow. (b) xy-averaged electrostatic potential profiles of the TTF-adsorbed phosphorene system under different external electric fields. The magnified potential profile around the TTF molecule are shown in the inset. Band structures of the TTF-adsorbed phosphorene system under the electric fields of (c) -0.5 V/Å and (d) 0.5 V/Å. Γ (0.0, 0.0, 0.0), X (0.0, 0.5, 0.0), M (0.5, 0.5, 0.0), and Y (0.5, 0.0, 0.0) refer to special points in the first Brillouin zone.

## IV. SUMMARY

In conclusion, on the basis of DFT calculations, we have performed a theoretical research on the structural and electronic properties of single-layer phosphorene functionalized by the organic molecules. Our calculations show that band gap of the adsorption complexes decreases with different organic molecules adsorbed because of the impurity states from the molecule. The typical p-type doped phosphorene can be obtained by adsorbing with the typical electrophilic molecule TCNQ which introduces the shallow acceptor states, with considerable electron transfer from phosphorene to TCNQ. While, when the typical nucleophilic molecule TTF is adsorbed on the phosphorene, the introduced donor states are far from the phosphorene CBE so that the adsorption complex does not become effectively n-doped and the corresponding charge transfer is also very small. But when an appropriate external electric field perpendicular to the phosphorene surface with direction from phosphorene to TTF is applied, the deep donor states will move closer to the phosphorene CBE and turn to the shallow ones, so an effectively n-doped phosphorene can be obtained. Our results suggest the electronic structure of phosphorene can be controlled over a wider range by the adsorption of organic molecules without destroying the phosphorene structural integrity. Thus, it is possible to build phosphorene-based electronic devices by modifying the phosphorene with patterned organic molecules, which may be broaden the application of phosphorene in the rising nanoelectronics.

**AUTHOR INFORMATION**

**Corresponding Author**

\* E-mail: jlyang@ustc.edu.cn. Phone: +86-551-63606408. Fax: +86-551-63603748 (J. Y.).

**Author Contributions**

The manuscript was written through contributions of all authors. All authors have given approval to the final version of the manuscript.

**ACKNOWLEDGMENT**

This work is partially supported by the National Key Basic Research Program (2011CB921404), by NSFC (21121003, 91021004, 21233007, 21273210), by CAS (XDB01020300), by Fundamental Research Funds for the Central Universities, and by USTCSCC, SCCAS, Tianjin, and Shanghai Supercomputer Centers.